\documentstyle[12pt]{article}
\def\be{\begin{equation}}
\def\ee{\end{equation}}
\def\bea{\begin{eqnarray}}
\def\eea{\end{eqnarray}}

\begin{document}

\begin{flushright}
IPM/P-2001/002\\
\end{flushright}

\begin{center}
{\Large{\bf T-duality and Noncommutative DBI Action}}

\vskip .5cm
{\large Davoud Kamani}
\vskip .1cm
 {\it Institute for Studies in Theoretical Physics and
Mathematics (IPM)
\\  P.O.Box: 19395-5531, Tehran, Iran}\\
{\sl e-mail: kamani@theory.ipm.ac.ir}
\\
\end{center}

\begin{abstract}

In this article we study the noncommutative description of
the DBI Lagrangian and its T-dual counterpart. 
We restrict the freedoms of the noncommutativity
parameters of these Lagrangians. 
Therefore the noncommutativity parameter, the effective metric, the effective
coupling constant of the string and the extra modulus ${\tilde \Phi}$ 
of the effective T-dual theory, can be expressed in
terms of the closed string variables $g$, $B$, $g_s$ and the 
noncommutativity parameter of the effective theory of open string.

\end{abstract}
\vskip .5cm

\newpage
%%%%%%%%%%%%%%%%%%%%%%%%%%%%%%%%%%%%%%%%%%%%%%%%%%%%%%%%%%%%%%%%%%%%%%%%%%%%%
\section{Introduction}
Over the past years there have been much activities exploring the 
relation between string theory and noncommutative geometry \cite{1,2,3,4}.
There have been attempts to explain noncommutativity on D-brane worldvolume
through the study of open strings in the presence of background fields
\cite{3,4}. For slowly varying fields, 
the effective Lagrangian of the open string
theory is the Dirac-Born-Infeld (DBI) Lagrangian \cite{5}.
The equivalence of noncommutative and ordinary DBI theory was 
proven \cite{1}. 
There are some general descriptions for the open string effective action
such that noncommutativity parameter is arbitrary. In other words,
a general DBI theory has been proposed to be described by a noncommutative
action including an extra modulus $\Phi$ \cite{1}.
The modules and their T-dualities have been considered
\cite{6,7}, and the relation to the DBI Lagrangian has been 
explored \cite{7}. Other aspects of the noncommutative DBI theory have been 
discussed in Ref.\cite{8}.

We observe that the open string metric and the noncommutativity parameter
appear as the background fields of the T-dual theory of string theory.
Also the background fields of string theory appear as the effective metric and 
the noncommutativity parameter of the T-dual theory of the effective theory.

We concentrate on the freedoms of the noncommutative DBI action and its 
T-dual counterpart. The freedoms of the noncommutativity parameters  
will be restricted. In this case the noncommutativity parameter,
the effective metric, 
the effective coupling constant of the string and the modulus
${\tilde \Phi}$ of the effective T-dual theory 
are expressed in terms of the closed string 
variables $g$, $B$, $g_s$ and the  
noncommutativity parameter of the effective theory of open string.

This paper is organized as follows. In section 2, we
give a brief review of the T-duality of 
the string action with the background
fields $g$ and $B$. In section 3, the noncommutative descriptions of the open
string effective action and its T-dual theory will be
discussed. In section 4, relations between the effective variables of the 
noncommutative DBI Lagrangian and the variables of its T-dual counterpart
will be discussed.
%%%%%%%%%%%%%%%%%%%%%%%%%%%%%%%%%%%%%%%%%%%%%%%%%%%%%%%%%%%%%%%%%%%%%%%%%%%%%%
\section{String action and its T-duality }

A fundamental bosonic string ending on a D$_p$-brane
with the background fields $g$ and $B$ has the action \cite{9}
\bea
S=\frac{1}{4\pi \alpha'} \int_{\Sigma} d^2 \sigma
\bigg{(} g_{ij} \partial_a
X^i \partial^a X^j -2\pi i\alpha' \epsilon^{ab} B_{ij}
\partial_a X^i \partial_b X^j \bigg{)}\;,
\eea
where $\Sigma$ is the string worldsheet. Just like
Ref.\cite{1}, we assume that the target space and the string worldsheet 
are Euclidean, and 
$B_{ij}$ is non-zero only for $i,j=1,...,r$ and $g_{ij}$ vanishes
for $i=1,...,r$ , 
$j \neq 1,...,r$ , where $r \leq p+1$ .
Further more assume $g_{ij}$ and $B_{ij}$ to be constant background fields. 

Now consider the T-duality of the theory. In the case of toroidal
compactification, when $d$-spatial coordinates are compactified on 
the torus $T^d$, the T-duality group is $O(d,d; {\bf Z})$ \cite{10}. Assume
that the D$_p$-brane is wrapped on the torus $T^p$. 
A particular element of $O(p,p;{\bf Z})$ T-duality group is
\bea
T= \left( \begin{array}{cc}
{\bf 0} & {\bf 1}_{p \times p}\\
{\bf 1}_{p \times p} & {\bf 0}
\end{array} \right).
\eea
Under the action of this element of the T-duality group, we have  
the following transformation for the background fields $g$ and $B$ \cite{11}
\bea
(g+2\pi \alpha' B) \rightarrow ( {\tilde g}+ 2\pi
\alpha' {\tilde B})=
(g+2\pi \alpha' B)^{-1}\;.
\eea
In other words, T-duality transformations of $g$ and $B$ are 
\bea
&~& {\tilde g} = (g+2 \pi \alpha' B)^{-1} g (g-2 \pi \alpha' B)^{-1}\;,
\\
&~& {\tilde B} = -(g+2 \pi \alpha' B)^{-1} B (g-2 \pi
\alpha' B)^{-1}\;,
\eea
where ${\tilde g}$ is symmetric and ${\tilde B}$ is antisymmetric. 
These are background fields of the T-dual theory. Also the
string coupling $g_s$ transforms as
\bea
{\tilde g}_s = \frac{g_s}{\sqrt{\det(g+2\pi \alpha'B)}}\;.
\eea

The action of the dual theory can be written as the following
\bea
{\tilde S}=\frac{1}{4\pi \alpha'} \int d^2 \sigma \bigg{(}
{\tilde g}_{ij} \partial_a {\tilde X}^i \partial^a
{\tilde X}^j -2\pi i\alpha'
\epsilon^{ab} {\tilde B}_{ij} \partial_a {\tilde X}^i
\partial_b {\tilde X}^j \bigg{)}\;. 
\eea
The equation of motion and the boundary conditions 
resulted from this action imply that $\{ {\tilde X}^i \}$ are coordinates 
on a noncommutative space.
%%%%%%%%%%%%%%%%%%%%%%%%%%%%%%%%%%%%%%%%%%%%%%%%%%%%%%%%%%%%%%%%%%%%%%%%%%%
\section{Noncommutativity of the effective theories}

Now let us discuss the
noncommutative description of the DBI Lagrangian and the T-duality of it. 
The noncommutative description can be expressed in terms of
noncommutative gauge field and the open string
variables $G_{(0)}$ , $\theta_0$ and $G^{(0)}_s$ \cite{1}
\bea
{\widehat {\cal{L}}}_{(0)} = \frac{1}{(2\pi)^p
(\alpha')^{\frac{p+1}{2}}
G^{(0)}_s } \sqrt{\det ( G_{(0)}+ 2\pi \alpha' {\widehat F} )}\;,
\eea
where the open string variables are
\bea
&~& G^{ij}_{(0)} = \bigg{(}(g+2 \pi \alpha' B)^{-1} g (g-2
\pi \alpha' B)^{-1} \bigg{)}^{ij}\;,
\nonumber\\
&~& G_{(0)ij} = \bigg{(}(g-2 \pi \alpha' B)g^{-1} (g+2 \pi
\alpha' B) \bigg{)}_{ij}\;, 
\nonumber\\
&~& \theta^{ij}_0 = -(2 \pi \alpha')^2 \bigg{(}(g+2 \pi
\alpha' B)^{-1} B (g-2 \pi
\alpha' B)^{-1} \bigg{)}^{ij}\;,
\nonumber\\
&~& G^{(0)}_s = g_s \bigg{(} \frac{\det G_{(0)}}{\det
(g+2\pi \alpha'B)}\bigg{)}^{\frac{1}{2}} \;.
\eea
These equations and the equations (4) and (5) give the following results
\bea
&~& G_{(0)} = {\tilde g}\;,
\nonumber\\
&~& \theta_0 = (2 \pi \alpha' )^2 {\tilde B}\;,
\eea
i.e. the open string metric and the noncommutativity parameter appear as 
the background fields of the T-dual theory of string theory.

The T-dual description of the Lagrangian (8) can be obtained from (8) by 
the parameters ${\tilde G}^{(0)}_s,\;{\tilde G}_{(0)}$ 
and ${\tilde \theta}_0$. These parameters obey the equations 
(9) which the parameters $g_s$, $g$ and $B$ should be changed by the 
parameters ${\tilde g}_s,\;{\tilde g}$ and ${\tilde B}$ respectively. 
Therefore similar to the relations (10), we obtain 
\bea
&~& {\tilde G}_{(0)} = g\;,
\nonumber\\
&~& {\tilde \theta}_0 = (2\pi \alpha')^2 B\;,
\eea
that is the background fields of the string theory appear as the 
effective metric and the noncommutativity parameter of the 
effective theory of the T-dual theory.

There is a two-form $\Phi$ such that for the effective theory 
more general description with an arbitrary
parameter $\theta$ is possible \cite{1}. 
In this case the effective Lagrangian is 
\bea
{\widehat {\cal{L}}} = \frac{1}{(2\pi)^p
(\alpha')^{\frac{p+1}{2}} G_s }
\sqrt{\det \bigg{(} G+ 2\pi \alpha' (\Phi+{\widehat F}) \bigg{)}}\;,
\eea
where the variables $G$, $\Phi$ and $G_s$ depend on
$g$, $B$ and $\theta$,
\bea
\frac{1}{G+2\pi \alpha' \Phi} = - \frac{\theta}{2\pi 
\alpha'}+\frac{1}{g+2\pi \alpha' B}\;,
\eea
\bea
G_s = g_s \bigg{(} \frac{\det( G+2\pi \alpha'
\Phi)}{\det (g+2\pi \alpha'B)}\bigg{)}^{\frac{1}{2}} \;.
\eea
Similarly there is such freedom for the effective
Lagrangian of the dual theory. 

Since the noncommutativity parameter of the T-dual theory of (12)
i.e. ${\tilde \theta}$
is independent of the variables of the original theory, i.e. $g$, $B$ and
$g_s$, and also it is independent of the parameter $\theta$, the  
T-dual theory of (12) partially 
depends on the original theory and its effective
theory, but not completely. We shall introduce an appropriate 
and consistent relation such that the partially
dependence changes to the completely dependence. That is, ${\tilde \theta}$
will be expressed in terms of $g$, $B$ and $\theta$.
%%%%%%%%%%%%%%%%%%%%%%%%%%%%%%%%%%%%%%%%%%%%%%%%%%%%%%%%%%%%%%%%%%%%%%%%%%%%%%%%
\section{Relations between the effective variables}

In the effective action and in the T-dual theory of it, there are 
two arbitrary noncommutativity parameters $\theta$ and ${\tilde \theta}$.
These parameters appear
in the variables $G$, $\Phi$, ${\tilde G}$ and
${\tilde \Phi}$ and also in the
effective string couplings $G_s$ and ${\tilde G}_s$.
We can express one of them
in terms of the other. For this, we introduce the following relation
\bea
{\tilde G}+2\pi \alpha' {\tilde \Phi}=(g+2\pi 
\alpha'B)^{-1}(G-2\pi\alpha' \Phi)(g-2\pi\alpha'B)^{-1}\;.
\eea
From the equation (13) and its T-dual analogue and the equation (15) 
we obtain 
\bea
{\tilde \theta}=-(g-2\pi \alpha'B) \theta (g+2\pi\alpha'B)\;,
\eea
which implies, if the effective theory is noncommutative (ordinary) the 
effective theory of the T-dual theory also 
is noncommutative (ordinary) and vice-versa. Equation
(15) has interesting properties that we present some of them.

As expected, equation (15) also holds under the following exchanges
\bea
G \leftrightarrow {\tilde G}\;,
\nonumber\\
\Phi \leftrightarrow {\tilde \Phi}\;,
\nonumber\\
g \rightarrow {\tilde g}\;,
\nonumber\\
B \rightarrow {\tilde B}\;.
\eea
To see this property, perform the transposed of both sides of the
equation (15) and then use the equation (3).

Consider the commutative case i.e. $\theta = 0$. Equation (16) also gives 
${\tilde \theta}=0$. Therefore there are 
$G=g$, $\Phi = B$, ${\tilde G}={\tilde 
g}$ and ${\tilde \Phi}={\tilde B}$. According to these, the
equation (15) reduces to the duality equation (3), which is an expected
result.

The following values for the variables $\theta$, $G$
and $\Phi$ (for the maximal rank $r=p+1$)
\bea
&~& \theta = B^{-1}\;,
\nonumber\\
&~& G = -(2\pi\alpha')^2 Bg^{-1}B\;,
\nonumber\\
&~& \Phi = -B\;,
\eea
satisfy the equation (13) \cite{1}. 
Also consider the dual variables with the forms
\bea
&~& {\tilde \theta} = {\tilde B}^{-1}
=-(g-2\pi \alpha'B) B^{-1}(g+2\pi\alpha'B)\;, 
\nonumber\\
&~& {\tilde G} = -(2\pi\alpha')^2 {\tilde B}{\tilde g}^{-1}{\tilde B}=
-(2\pi \alpha')^2(g+2\pi\alpha'B)^{-1} Bg^{-1}B (g-2\pi\alpha'B)^{-1} \;,
\nonumber\\ 
&~& {\tilde \Phi} = -{\tilde B} =(g+2\pi
\alpha'B)^{-1}B (g-2\pi\alpha'B)^{-1}\;, 
\eea
where ${\tilde g}$ and ${\tilde B}$ in terms of $g$
and $B$ are used. These parameters satisfy the T-dual analogue of 
the equation (13). 
Therefore for the solutions (18) and (19), the
equation (15) reduces to an identity.

According to the equation (15), we can express the dual
variables ${\tilde G}$, ${\tilde \Phi}$ and ${\tilde G}_s$ in terms of 
the parameters $g$, $B$, $\theta$ and $g_s$. Using the fact that for a 
matrix $M$ there is the identity $\det M = \det M^T$, we obtain
\bea
{\tilde G}_s = \frac{{\tilde g_s}}{g_s}G_s =
\frac{G_s}{\sqrt{\det (g+2\pi \alpha'B)}}\;,
\eea
which implies that the quantity $\frac{G_s}{g_s}$ is T-duality
invariant. In other words $\frac{g_{YM}}{\sqrt{g_s}}$ is invariant, i.e.
\bea
{\tilde g}_{YM} = g_{YM} \sqrt{\frac{{\tilde g}_s}{g_s}}=g_{YM}
[\det(g+2\pi \alpha'B)]^{-\frac{1}{4}}\;.
\eea
The dual variables ${\tilde G}$ and ${\tilde \Phi}$
are symmetric and antisymmetric parts of the right hand side of the
equation (15), i.e.
\bea
&~& {\tilde G} = (g+2 \pi \alpha' B)^{-1} G (g-2 \pi \alpha' B)^{-1}\;,
\\
&~& {\tilde \Phi} = -(g+2 \pi \alpha' B)^{-1} \Phi (g-2 \pi \alpha' B)^{-1}\;,
\eea
where $G$ and $\Phi$ are given by the equation (13).
The equations (22) and (23) for $\theta=0$, reduce to the
results ${\tilde G} ={\tilde g}$ and ${\tilde \Phi} = {\tilde B}$. They 
also respect the exchanges (17), and for the solutions (18)
and (19) change to identities.

Consider the interesting special case $\theta = \theta_0 $. From the 
equation (16) we obtain
\bea
{\tilde \theta}={\tilde \theta}_0=(2\pi\alpha')^2 B\;.
\eea
Therefore the antisymmetric tensors $\Phi$ and ${\tilde \Phi}$ vanish
which means, if the freedom of the parameter $\theta$ removes, the 
freedom in the T-dual theory also removes. For this special case the
effective metric ${\tilde G}$ and 
the effective coupling constant ${\tilde G}_s$ reduce to
\bea
&~& {\tilde G} = {\tilde G}_{(0)}\;,
\nonumber\\
&~& {\tilde G}_s = {\tilde G^{(0)}}_s = \frac{g_s}{\sqrt{\det g}}\;.
\eea

Note that according to the equations (20) and (22) the quantity
$\frac{\sqrt{\det G}}{G^2_s}$ is T-duality invariant. 
%%%%%%%%%%%%%%%%%%%%%%%%%%%%%%%%%%%%%%%%%%%%%%%%%%%%%%%%%%%%%%%%%%%%%%%%%%%
\section{Conclusion}

We saw that when the extra modulus $\Phi$ is zero (i.e.
$\theta = \theta_0$), the background fields of string theory
(T-dual of string theory), are the 
effective metric and the noncommutativity 
parameter of the effective T-dual theory 
(the effective theory of string theory).

We introduced a consistent relation between  
the open string variables $( G , \Phi )$
and the dual variables $({\tilde G}, {\tilde \Phi})$.
Therefore the noncommutativity parameters are expressed 
in terms of each other. In 
this case if the effective theory is noncommutative (ordinary), 
the effective T-dual 
theory also is noncommutative (ordinary) and vice-versa. The ratio of
the effective open string coupling to the open string 
coupling is a T-duality invariant quantity. In the effective theory of the 
T-dual theory the noncommutativity parameter, 
the effective metric, the effective coupling of string
and the antisymmetric tensor ${\tilde \Phi}$ 
are expressed in terms of the closed string variables 
$g$, $B$, $g_s$ and the noncommutativity parameter $\theta$.

{\bf Acknowledgement}

The author would like to thank H. Arfaei for useful discussion.
%%%%%%%%%%%%%%%%%%%%%%%%%%%%%%%%%%%%%%%%%%%%%%%%%%%%%%%%%%%%%%%%%%%%%%%%%%%

\end{document}